\documentclass[12pt,a4paper]{article}
\usepackage{a4wide}
\usepackage{amsmath}
\usepackage{amssymb}
\usepackage{amsfonts}
\usepackage{epsfig}
\usepackage{subfigure}
\usepackage{exscale}
\usepackage{float}
\usepackage{bbm}
\usepackage[numbers,sort&compress]{natbib}

\newcommand{\Z}{{\mathbb{Z}}}

\newcommand{\C}{{\mathbb{C}}}

\newcommand{\1}{{\mathbbm{1}}}
\newcommand{\p}{\partial}

\makeatletter
\@addtoreset{equation}{section}
\makeatother

\title{Discrete Accidental Symmetry for a Particle in a Constant Magnetic Field
on a Torus}

\author{M.\ H.\ Al-Hashimi and U.-J.\ Wiese \\ \\
Institute for Theoretical Physics, Bern University \\
Sidlerstrasse 5, CH-3012 Bern, Switzerland \\ \\}

\begin{document} 

\maketitle

\vspace{-1cm}

\begin{abstract} \normalsize

A classical particle in a constant magnetic field undergoes cyclotron motion on
a circular orbit. At the quantum level, the fact that all classical orbits are
closed gives rise to degeneracies in the spectrum. It is well-known that the 
spectrum of a charged particle in a constant magnetic field consists of 
infinitely degenerate Landau levels. Just as for the $1/r$ and $r^2$ 
potentials, one thus expects some hidden accidental symmetry, in this case with
infinite-dimensional representations. Indeed, the position of the center of the
cyclotron circle plays the role of a Runge-Lenz vector. After identifying the 
corresponding accidental symmetry algebra, we re-analyze the system in a finite
periodic volume. Interestingly, similar to the quantum mechanical breaking of
CP invariance due to the $\theta$-vacuum angle in non-Abelian gauge theories, 
quantum effects due to two self-adjoint extension parameters $\theta_x$ and
$\theta_y$ explicitly break the continuous translation invariance of the 
classical theory. This reduces the symmetry to a discrete magnetic translation 
group and leads to finite degeneracy. Similar to a particle moving on a cone, a
particle in a constant magnetic field shows a very peculiar realization of 
accidental symmetry in quantum mechanics.

\end{abstract}

\newpage
 
\section{Introduction}

The fact that for some physical systems all bound classical orbits are closed 
leads to accidental degeneracies in the discrete energy spectrum of the
corresponding quantum systems. Accidental symmetries are familiar from a
particle moving in a $1/r$ or $r^2$ potential. In $d$ spatial dimensions the
system then has an $SO(d)$ rotational symmetry. In case of the $1/r$ potential,
this symmetry is dynamically enhanced to an accidental $SO(d+1)$ symmetry, and 
for the $r^2$ harmonic oscillator potential it is enhanced to $SU(d)$. The
accidental symmetries give rise to additional degeneracies in the discrete
energy spectrum of the corresponding quantum systems, beyond the degeneracies
one would expect based on rotation invariance alone \cite{Foc35,Bar36}. The 
components of the
Runge-Lenz vector \cite{Len24} are the generators of the accidental symmetry 
algebras.  The subject of accidental symmetry has been reviewed, for example, 
by McIntosh \cite{McI71}. Recently, we have further investigated the phenomenon
of accidental symmetries, by studying a particle confined to the surface of a
cone and bound to its tip by a $1/r$ or $r^2$ potential \cite{Has07}. When the
deficit angle of the cone is a rational fraction of $2 \pi$, again all bound
classical orbits are closed and there are accidental degeneracies in the energy
spectrum of the quantum system. In this case the Runge-Lenz vector does not act
as a self-adjoint operator in the domain of the Hamiltonian. Remarkably, as a 
consequence of this unusual property, the accidental $SU(2)$ symmetry has 
unusual multiplets with fractional (i.e.\ neither integer nor half-integer)
spin.

An interesting example of an accidental symmetry 
involving a vector potential is cyclotron motion \cite{Lan30,Joh49}. Also in 
this case, there is a deep connection between the fact that all bound classical
orbits are closed and additional degeneracies in the discrete energy spectrum 
of the corresponding quantum system. As was already noted in \cite{Joh49}, the 
center of the circular cyclotron orbit is a conserved quantity analogous to the
Runge-Lenz vector in the Kepler problem. Also the radius of the cyclotron orbit
is a conserved quantity directly related to the energy. Interestingly, while 
the two coordinates of the center are not simultaneously measurable, the 
radius of the circle has a sharp value in an energy eigenstate. In the 
cyclotron problem, translation invariance disguises itself as an ``accidental''
symmetry. As a consequence, the symmetry multiplets --- i.e.\ the Landau levels
--- are infinitely degenerate. In order to further investigate the nature of 
the accidental symmetry, in \cite{Dul66} the charged particle in the magnetic 
field was coupled to the origin by an $r^2$ harmonic oscillator potential. This
explicitly breaks translation invariance and thus reduces the degeneracy to a 
finite amount, while rotation invariance remains intact. In this paper, we do
the opposite, i.e.\ we explicitly break rotation invariance, while leaving 
translation invariance (and hence the accidental symmetry) intact by putting 
the system on a torus. Interestingly, the Polyakov loops, which are a 
consequence of the non-trivial holonomies of the torus, give rise to 
non-trivial Aharonov-Bohm phases which are observable at the quantum but not at
the classical level. Analogous to the quantum mechanical breaking of CP 
invariance due to the $\theta$-vacuum angle in non-Abelian gauge theories, here
two self-adjoint extension parameters $\theta_x$ and $\theta_y$ explicitly 
break the continuous translation invariance of the classical problem down to a 
discrete magnetic translation group \cite{Zak64}. This reduces the degeneracy 
to a finite 
amount, and allows us to further investigate the nature of the accidental 
symmetry. In particular, just like for motion on a cone \cite{Has07}, symmetry 
manifests itself in a rather unusual way in this quantum system. In particular,
due to its relevance to the quantum Hall effect, the Landau level problem has 
been studied very extensively (for a recent review see \cite{Ste08}). For 
example, the problem has already been investigated on a torus in 
\cite{Che95,Zai89}, however, without emphasizing the accidental symmetry 
aspects. In this paper, we concentrate entirely on those aspects, thus 
addressing an old and rather well-studied problem from an unconventional point 
of view.

The rest of the paper is organized as follows. In section 2 the cyclotron 
problem is reviewed in the infinite volume, with special emphasis on its
oscillator algebras and accidental symmetry generators. In section 3 the 
system is put on a torus and the unusual manifestation of the accidental
symmetry is worked out. Section 4 contains our conclusions.

\section{Particle in the Infinite Volume}

In this section we review the standard knowledge about a non-relativistic 
particle moving in a constant magnetic field in the infinite volume. We proceed
from a classical to a semi-classical, and finally to a fully quantum mechanical
treatment. In particular, we emphasize the symmetry aspects of the problem with
a focus on accidental symmetries. This section is a preparation for the case of
a finite periodic volume to be discussed in the next section. In the following,
we will use natural units in which $\hbar = c = 1$.

\subsection{Classical Treatment}

Ignoring its spin, we consider a non-relativistic electron of mass $M$ and 
electric charge $-e$ moving in a constant magnetic field $\vec B = B \vec e_z$,
which we realize through the vector potential
\begin{equation}
A_x(\vec x) = 0, \ A_y(\vec x) = B x, \ A_z(\vec x) = 0.
\end{equation} 
Since the motion along the direction of the magnetic field is trivial, we 
restrict ourselves to 2-dimensional motion in the $x$-$y$-plane. Obviously, 
this is just standard cyclotron motion. To get started, in this subsection we 
treat the problem classically. The particle then experiences the Lorentz force
\begin{equation}
\vec F(t) = - e \vec v(t) \times B \vec e_z,
\end{equation}
which forces the particle on a circular orbit of some radius $r$. It moves 
along the circle with an angular velocity $\omega$, which implies the linear
velocity $v = \omega r$ and the acceleration $a = \omega^2 r$. Hence, Newton's
equation takes the form
\begin{equation}
m \omega^2 r = e \omega r B \ \Rightarrow \ \omega = \frac{e B}{M},
\end{equation}
with the cyclotron frequency $\omega$ being independent of the radius $r$.
Obviously, for this system all classical orbits are closed. The same is true
for a particle moving in a $1/r$ or $r^2$ potential. In those cases, the fact
that all bound classical orbits are closed is related to the conservation of 
the Runge-Lenz vector which generates a hidden accidental dynamical symmetry. 

Let us now investigate the question of accidental symmetry for the particle in 
the constant magnetic field. The Lagrange function then takes the form
\begin{equation}
L = \frac{M}{2} \vec v^2 - e \vec A(\vec x) \cdot \vec v  =
\frac{M}{2} \left(\dot x^2 + \dot y^2\right) - e B x \dot y,
\end{equation}
and the corresponding conjugate momenta are
\begin{equation}
p_x = \frac{\p L}{\p \dot x} = M \dot x = M v_x, \
p_y = \frac{\p L}{\p \dot y} = M \dot y - e B x = M v_x - e B x.
\end{equation}
First of all, in the gauge that we picked, $y$ is a cyclic coordinate and hence
the canonically conjugate momentum $p_y$ is conserved as a consequence of
translation invariance in the $y$-direction. Despite the fact that the system 
is translation invariant also in the $x$-direction, $x$ itself is not a cyclic
coordinate and hence $p_x$ is not conserved. Still, using Noether's theorem one
can identify the corresponding conserved quantity as $P_x = p_x + e B y$.
Interestingly, the Lagrange function is not invariant under a shift in the
$x$-direction but changes by a total derivative (which leaves the classical
equations of motion unchanged).

The classical Hamilton function takes the form
\begin{equation}
\label{Hamiltonian}
H = \vec p \cdot \vec v - L = 
\frac{1}{2 M}\left[\vec p + e \vec A(\vec x)\right]^2 =
\frac{1}{2 M}\left[p_x^2 + \left(p_y + e B x\right)^2\right].
\end{equation}
It is straightforward to convince oneself that $H$ has vanishing Poisson
brackets, $\{H,P_x\} = \{H,P_y\} = \{H,L\} = 0$, with the three symmetry 
generators
\begin{equation}
P_x = p_x + e B y, \ P_y = p_y, \
L = x \left(p_y + \frac{e B}{2} x\right) - 
y \left(p_x + \frac{e B}{2} y\right).
\end{equation}
One can identify $P_x$, $P_y$, and $L$ as the gauge-covariant 
generators of translations and rotations. In particular, one obtains
\begin{equation}
\{L,P_x\} = P_y, \ \{L,P_y\} = - P_x,
\end{equation}
as one would expect for the rotation properties of the vector $(P_x,P_y)$. As
is well-known, however, in a magnetic field the two translations $P_x$ and
$P_y$ do not commute, i.e.
\begin{equation}
\{P_x,P_y\} = e B.
\end{equation}
How can these standard symmetry considerations be related to an accidental
symmetry due to a Runge-Lenz vector? The Runge-Lenz vector is familiar from
the Kepler problem. It points from the center of force to the perihelion
position, and is conserved because all bound classical orbits are closed.
Similarly, the orbit of a charged particle in a constant magnetic field is a 
closed circle with a fixed center. Indeed, in this case the position of this 
center plays the role of the conserved Runge-Lenz vector and is given by
\begin{eqnarray}
\label{RP}
&&R_x = x - \frac{v_y}{v} r = x - \frac{v_y}{\omega} = 
x - \frac{1}{M \omega}\left(p_y + e B x\right) = 
- \frac{p_y}{M \omega} = - \frac{P_y}{e B}, \nonumber \\
&&R_y = y + \frac{v_x}{v} r = y + \frac{v_x}{\omega} = 
y + \frac{p_x}{M \omega} = \frac{P_x}{e B}.
\end{eqnarray}
Interestingly, the position $(R_x,R_y)$ of the center of the cyclotron circle 
is, at the same time, proportional to $(- P_y,P_x)$, i.e.\ it is orthogonal to
the generators of spatial translations. Consequently, we can write
\begin{equation}
\label{poisson1}
\{R_x,P_x\} = - \frac{1}{e B} \{P_y,P_x\} = 1, \
\{R_y,P_y\} = \frac{1}{e B} \{P_x,P_y\} = 1, \
\end{equation}
as well as
\begin{equation}
\label{poisson2}
\{R_x,P_y\} = - \frac{1}{e B} \{P_y,P_y\} = 0, \
\{R_y,P_x\} = \frac{1}{e B} \{P_x,P_x\} = 0.
\end{equation}
While eqs.(\ref{poisson1}) and (\ref{poisson2}) look like the usual Poisson 
brackets of position and momentum, one should not forget that $R_x$ and $R_y$ 
are just multiples of $P_y$ and $P_x$, and should hence not be mistaken as
independent variables. In particular, one also obtains the relation
\begin{equation}
\{R_x,R_y\} = \frac{1}{e B}.
\end{equation}
Hence, just like the two generators of translations, the $x$- and 
$y$-components of the Runge-Lenz vector do not have a vanishing Poisson
bracket. At the quantum level, this will imply that the $x$- and $y$-components
of the center of a cyclotron circle are not simultaneously measurable with 
absolute precision.

Another conserved quantity is the radius $r$ of the circular cyclotron orbit
which can be expressed as
\begin{equation}
r^2 = (x - R_x)^2 + (y - R_y)^2 = \frac{1}{M^2 \omega^2} 
\left(p_y + e B x\right)^2 + \frac{p_x^2}{M^2 \omega^2} = 
\frac{2 H}{M \omega^2}.
\end{equation}
Since $r^2$ is proportional to the energy, it obviously is indeed conserved.

\subsection{Semi-classical Treatment}

Next, we consider the same problem semi-classically, i.e.\ by using 
Bohr-Sommerfeld quantization, which, in this case, is equivalent to the 
quantization of angular momentum, i.e.\ $L = n$. For a cyclotron orbit of 
radius $r$, it is easy to convince oneself that
\begin{equation}
L = \frac{e B}{2} r^2 = n \ \Rightarrow \ r = \sqrt{\frac{2 n}{e B}}.
\end{equation}
Consequently, in the semi-classical treatment the allowed radii of cyclotron 
orbits are now quantized. Using eq.(\ref{Hamiltonian}) one finds for the energy
\begin{equation}
E = H = \frac{1}{2} M \omega^2 r^2 = n \omega.
\end{equation}
As is well-known, up to a constant $\frac{\omega}{2}$, the semi-classically 
quantized energy values are those of a harmonic oscillator with the cyclotron
frequency $\omega$.

\subsection{Quantum Mechanical Treatment}

Finally, we consider the problem fully quantum mechanically. The Schr\"odinger 
equation then takes the form
\begin{equation}
- \frac{1}{2 M} \left[\partial_x^2 + 
\left(\partial_y + i e B x\right)^2\right] \Psi(\vec x) =
E \Psi(\vec x).
\end{equation}
We now make the factorization ansatz
\begin{equation}
\label{factorization}
\Psi(\vec x) = \psi(x) \exp(i p_y y),
\end{equation}
and we obtain
\begin{equation}
\left[- \frac{\p_x^2}{2 M} +
\frac{1}{2} M \omega^2 \left(x + \frac{p_y}{M \omega}\right)^2\right] 
\psi(x) = E \psi(x).
\end{equation}
Indeed, this is the Schr\"odinger equation of a shifted harmonic oscillator. 
Hence, the quantum mechanical energy spectrum takes the form
\begin{equation}
E = \omega \left(n + \frac{1}{2}\right).
\end{equation}
Interestingly, the energy of the charged particle is completely independent of
the transverse momentum $p_y$. As a result, the quantized Landau levels have a
continuous infinite degeneracy. The energy eigenstates are shifted
one-dimensional harmonic oscillator wave functions $\psi_n(x)$, i.e.
\begin{equation}
\label{eigenstates}
\langle \vec x|n p_y \rangle = \psi_n\left(x + \frac{p_y}{M \omega} \right)
\exp(i p_y y).
\end{equation}
Similarly, one can construct eigenstates of the generator 
$P_x = - i \p_x + e B y$ of infinitesimal translations (up to a gauge 
transformation) in the $x$-direction 
\begin{equation}
\label{eigenx}
\langle \vec x|n p_x \rangle = \psi_n\left(y - \frac{p_x}{M \omega} \right)
\exp(i p_x x) \exp(- i e B x y).
\end{equation}  
It is straightforward to show that the two sets of eigenstates 
$\langle \vec x|n p_y\rangle$ and $\langle \vec x|n p_x\rangle$ span the same 
subspace of localized states in the Hilbert space. 

Since all classical orbits are closed and the center of the cyclotron orbit
plays the role of a Runge-Lenz vector, it is natural to ask whether the 
degeneracy is caused by an accidental symmetry. Of course, since the Runge-Lenz
vector plays a dual role and is also generating translations (up to gauge
transformations), in this case the ``accidental'' symmetry would just be 
translation invariance. Indeed, in complete analogy to the classical case, it 
is easy to convince oneself that $[H,R_x] = [H,R_y] = [H,L] = 0$, with the 
Runge-Lenz vector and the angular momentum operator given by
\begin{eqnarray}
&&R_x = - \frac{P_y}{e B} = \frac{i \p_y}{e B}, \ 
R_y = \frac{P_x}{e B} = y - \frac{i \p_x}{e B}, 
\nonumber \\
&&L = x \left(- i \p_y + \frac{e B x}{2}\right) - 
y \left(- i \p_x + \frac{e B y}{2}\right).
\end{eqnarray}
As in the classical case, the radius of the cyclotron orbit squared is given by
\begin{equation}
r^2 = (x - R_x)^2 + (y - R_y)^2 = \left(x - \frac{i \p_y}{e B}\right)^2
- \frac{ \p_x^2}{e^2 B^2} = \frac{2 H}{M \omega^2},
\end{equation}
and is thus again a conserved quantity. In particular, we can express the
Hamiltonian as
\begin{equation}
H = \frac{1}{2} M \omega^2 r^2.
\end{equation}
Remarkably, although the two coordinates $R_x$ and $R_y$ of the center of the
cyclotron circle are not simultaneously measurable, its radius $r$ has a 
definite value in an energy eigenstate.

As it should, under spatial rotations the Runge-Lenz vector $(R_x,R_y)$ indeed
transforms as a vector, i.e.
\begin{equation}
[L,R_x] = i R_y, \ [L,R_y] = - i R_x.
\end{equation}
These relations suggest to introduce
\begin{equation}
R_\pm = R_x \pm i R_y,
\end{equation}
which implies
\begin{equation}
[L,R_\pm] = \pm R_\pm.
\end{equation}
Hence, $R_+$ and $R_-$ act as raising and lowering operators of angular
momentum. Still, it is important to note that $R_x$, $R_y$, and $L$ do not form
an $SU(2)$ algebra. This follows because, in analogy to the classical case
\begin{equation}
[R_x,R_y] = \frac{i}{e B},
\end{equation}
i.e.\ $R_x$ and $R_y$ are generators of a Heisenberg algebra. As a consequence
one obtains
\begin{equation}
[R_+,R_-] = \frac{2}{e B}.
\end{equation}

\subsection{Creation and Annihilation Operators}

Since the particle in the magnetic field leads to the spectrum of a 
1-dimensional harmonic oscillator (however, with infinite degeneracy), it is 
natural to ask how one can construct corresponding creation and annihilation 
operators such that
\begin{equation}
H = \omega \left(a^\dagger a + \frac{1}{2}\right), \ [a,a^\dagger] = 1.
\end{equation}
Remarkably, the creation and annihilation operators are closely related to the 
Runge-Lenz vector, i.e.\ the vector that points to the center of the 
classical cyclotron orbit. Since we have seen that
\begin{equation}
H = \frac{1}{2} M \omega^2 r^2 = 
\frac{1}{2} M \omega^2 \left[(x - R_x)^2 + (y - R_y)^2\right],
\end{equation}
one is led to identify
\begin{equation}
a = \sqrt{\frac{M \omega}{2}} \left[x - R_x - i (y - R_y)\right], \
a^\dagger = \sqrt{\frac{M \omega}{2}} \left[x - R_x + i (y - R_y)\right],
\end{equation}
which indeed have the desired properties. One also finds that
\begin{equation}
[L,a] = - a, \ [L,a^\dagger] = a^\dagger,
\end{equation}
which implies that $a^\dagger$ and $a$ also raise and lower the angular
momentum. Interestingly, we have seen before that
\begin{equation}
[L,R_\pm] = \pm R_\pm, \ [R_+,R_-] = \frac{2}{e B} = \frac{2}{M \omega}.
\end{equation}
Hence, $R_+$ and $R_-$ also act as raising and lowering operators of the 
angular momentum. Indeed, we can identify another set of creation and 
annihilation operators
\begin{equation}
b = \sqrt{\frac{M \omega}{2}} R_+, \ b^\dagger = \sqrt{\frac{M \omega}{2}} R_-,
\end{equation}
which obey
\begin{equation}
[L,b] = b, \ [L,b^\dagger] = - b^\dagger.
\end{equation}
As a result, $b$ raises and $b^\dagger$ lowers the angular momentum by one unit.
It is straightforward to derive the commutation relations
\begin{equation}
[a,b] = [a^\dagger,b] = [a,b^\dagger] = [a^\dagger,b^\dagger] = 0, \ 
[b,b^\dagger] = 1.
\end{equation}
Interestingly, just like a 2-dimensional harmonic oscillator, the particle in 
a magnetic field is described by two sets of commuting creation and 
annihilation operators. However, in contrast to the 2-dimensional harmonic 
oscillator, the Hamiltonian of the particle in a magnetic field contains only
$a^\dagger a$, but not $b^\dagger b$.

\subsection{Alternative Representation of the Hamiltonian}

Interestingly, the Hamiltonian can also be expressed as
\begin{equation}
H = \frac{1}{2} M \omega^2 \left(R_x^2 + R_y^2\right) + \omega L =
\omega \left(b^\dagger b + \frac{1}{2} + L\right) = H_0 + \omega L.
\end{equation}
Here we have introduced the Hamiltonian of an ordinary 1-dimensional harmonic
oscillator
\begin{equation}
H_0 = \omega \left(b^\dagger b + \frac{1}{2}\right),
\end{equation}
and the angular momentum operator has been identified as
\begin{equation}
L = a^\dagger a - b^\dagger b.
\end{equation}
Interestingly, the creation and annihilation operators $b^\dagger$ and $b$ 
commute with the total energy $H$ because they raise (lower) $H_0$ by $\omega$,
while they lower (raise) $L$ by 1, such that indeed
\begin{equation}
[H,b] = [H_0,b] + \omega [L,b] = 0, \
[H,b^\dagger] = [H_0,b^\dagger] + \omega [L,b^\dagger] = 0.
\end{equation}

\subsection{Energy Spectrum and Energy Eigenstates}

Since the algebraic structure of the problem (but not the exact form of the
Hamiltonian) is the same as for the 2-dimensional harmonic oscillator, we can
construct the physical states accordingly. First of all, we construct a state
$|0 0\rangle$ that is annihilated by both $a$ and $b$, i.e.
\begin{equation}
a|0 0\rangle = b|0 0\rangle = 0.
\end{equation}
Then we define states
\begin{equation}
|n n'\rangle = \frac{\left(a^\dagger\right)^{n}}{\sqrt{n!}} 
\frac{\left(b^\dagger\right)^{n'}}{\sqrt{n'!}} |0 0\rangle,
\end{equation}
which are eigenstates of the total energy
\begin{equation}
H |n n'\rangle = \omega \left(n + \frac{1}{2}\right) |n n'\rangle,
\end{equation}
as well as of the angular momentum
\begin{equation}
L |n n'\rangle = (n - n') |n n'\rangle = m |n n'\rangle.
\end{equation}
It should be noted that the quantum number $n \in \{0,1,2,...\}$ (which 
determines the energy) is non-negative, while the quantum number 
$m = n - n' \in \Z$ (which determines the angular momentum) is an arbitrary
integer. The infinite degeneracy of the Landau levels is now obvious 
because states with the same $n$ but different values of $n'$ have the same 
energy. 

One may wonder why in subsection 2.3 we found an infinite degeneracy labeled by
the continuous momentum $p_y$ and now we only find a countable variety of 
degenerate states (labeled by the integer $m$). This apparent discrepancy is
due to the implicit consideration of two different Hilbert spaces. While the
states in the discrete variety labeled by $m$ are normalizable in the usual
sense, the continuous variety of plane wave states labeled by $p_y$ is 
normalized to $\delta$-functions and thus belongs to an extended Hilbert space.

It is remarkable that a quantum mechanical system containing just a single 
particle has an even infinitely degenerate ground state. The existence of
infinitely degenerate ground states is usually associated with the spontaneous 
breakdown of a continuous global symmetry in systems with infinitely many 
degrees of freedom. Does the infinite degeneracy of the single-particle Landau 
levels have anything to do with the spontaneous breakdown of translation
invariance? The usual breaking of a continuous global symmetry is associated 
with the occurrence of massless Goldstone bosons. For example, when translation
invariance is spontaneously broken by the formation of a crystal lattice, 
phonons arise as massless excitations. In the quantum mechanical system studied
here, there is no room for phonons because it has only a finite number of 
degrees of freedom. Indeed, the infinitely degenerate ground states are 
separated from the rest of the spectrum by a gap $\omega$. Still, just like
a system with spontaneous symmetry breaking, the charged particle in a magnetic
field may chose spontaneously from a continuous variety of degenerate ground
states. 

\subsection{Coherent States}

Coherent states are well-known from the harmonic oscillator, and have also
been constructed for the Landau level problem \cite{Fel70}. As usual, the 
coherent states are constructed as eigenstates of the annihilation operators,
i.e.
\begin{equation}
a |\lambda \lambda'\rangle = \lambda |\lambda \lambda'\rangle, \
b |\lambda \lambda'\rangle = \lambda' |\lambda \lambda'\rangle, \ 
\lambda, \lambda' \in \C.
\end{equation}
In coordinate space, the coherent states can be expressed as
\begin{equation}
\langle \vec x|\lambda \lambda' \rangle = A
\exp\left[- \frac{M \omega}{4} (x^2 + 2 i x y + y^2) + 
\sqrt{\frac{M \omega}{2}}
\left(x (\lambda + \lambda') + i y (\lambda - \lambda')\right)\right].
\end{equation}
Some expectation values in the coherent state $|\lambda \lambda'\rangle$ are
given by
\begin{eqnarray}
&&\langle R_x \rangle = \sqrt{\frac{2}{M \omega}} \ \mbox{Re}\lambda', \
\Delta R_x = \frac{1}{\sqrt{2 M \omega}}, \nonumber \\
&&\langle R_y \rangle = \sqrt{\frac{2}{M \omega}} \ \mbox{Im}\lambda', \
\Delta R_y = \frac{1}{\sqrt{2 M \omega}}, \nonumber \\
&&\langle x - R_x \rangle = \sqrt{\frac{2}{M \omega}} \ \mbox{Re}\lambda, \
\Delta (x - R_x) = \frac{1}{\sqrt{2 M \omega}}, \nonumber \\
&&\langle y - R_y \rangle = - \sqrt{\frac{2}{M \omega}} \ \mbox{Im}\lambda, \
\Delta (y - R_y) = \frac{1}{\sqrt{2 M \omega}}, \nonumber \\
&&\langle M v_x \rangle =
\left\langle p_x + e A_x \right\rangle = 
\sqrt{2 M \omega} \ \mbox{Im}\lambda, \ 
\Delta(M v_x) = \sqrt{\frac{M \omega}{2}}, \nonumber \\
&&\langle M v_y \rangle =
\left\langle p_y + e A_y \right\rangle = 
\sqrt{2 M \omega} \ \mbox{Re}\lambda, \ 
\Delta(M v_y) = \sqrt{\frac{M \omega}{2}}, \nonumber \\
&&\langle H \rangle = \omega \left(|\lambda|^2 + \frac{1}{2}\right), \
\Delta H = \omega |\lambda|.
\end{eqnarray}
Here $\Delta O = \sqrt{\langle O^2 \rangle - \langle O \rangle^2}$ describes
the quantum uncertainty. In all cases $\Delta O/\langle O \rangle$ is
proportional to $1/|\lambda|$ or $1/|\lambda'|$, which implies that the 
relative uncertainty goes to zero in the classical limit.

Just as in the ordinary harmonic oscillator, the time-dependent Schr\"odinger
equation $i \p_t |\Psi(t)\rangle = H |\Psi(t)\rangle$ with an initial coherent
state $|\Psi(0)\rangle = |\lambda(0) \lambda' \rangle$ is (up to an
irrelevant phase) solved by $|\lambda(t) \lambda'\rangle$ with 
\begin{equation}
\lambda(t) = \lambda(0) \exp(- i \omega t).
\end{equation}
As expected, the state remains coherent during its time-evolution. In 
particular, this implies
\begin{eqnarray}
&&\langle x - R_x \rangle(t) = 
\frac{|\lambda|}{\sqrt{2 M \omega}} \cos(\omega t), \
\langle M v_x \rangle(t) = - \sqrt{2 M \omega} \ |\lambda| \sin(\omega t), 
\nonumber \\
&&\langle y - R_y \rangle(t) = 
\frac{|\lambda|}{\sqrt{2 M \omega}} \sin(\omega t), \
\langle M v_y \rangle(t) = \sqrt{2 M \omega} \ |\lambda| \cos(\omega t).
\end{eqnarray}
Hence, the coherent state represents a Gaussian wave packet moving around a
circular cyclotron orbit just like a classical particle. This is obvious from
the coordinate representation of the probability density
\begin{eqnarray}
&&|\langle \vec x|\Psi(t)\rangle|^2 = A \exp\left(- \frac{M \omega}{2}
\left[(x - \langle x \rangle(t))^2 + 
(y - \langle y \rangle(t))^2\right]\right), \nonumber \\
&&\langle x \rangle(t) = \langle R_x \rangle + 
\sqrt{\frac{2}{M \omega}} |\lambda| \cos(\omega t), \
\langle y \rangle(t) = \langle R_y \rangle + 
\sqrt{\frac{2}{M \omega}} |\lambda| \sin(\omega t).
\end{eqnarray}

It is interesting to note that the coherent states $|0 \lambda'\rangle$ with 
$\lambda = 0$ (but with arbitrary 
$\lambda' = \sqrt{M \omega/2} (\langle R_x \rangle + i \langle R_y \rangle)$ 
form an overcomplete set of degenerate ground states with the energy 
$\omega/2$. These states represent Gaussian wave packets 
centered at the points $(\langle R_x \rangle,\langle R_y \rangle)$ determined 
by $\lambda'$. Unlike for a free particle, these Gaussian wave packets do not
spread. Semi-classically speaking, the charged particle is in a ``circular 
orbit'' of quantized sharp radius $\sqrt{2/M \omega} |\lambda| = 0$ with an 
uncertain position $(\langle R_x \rangle,\langle R_y \rangle)$ of the center.
Since the ground state is infinite degenerate, the charged particle can 
spontaneously select any average position 
$(\langle R_x \rangle,\langle R_y \rangle)$ at which it can stay with average
velocity zero in a state of minimal uncertainty. Again, this is reminiscent of
the spontaneous breakdown of translation invariance.

\section{Particle on a Torus}

In this section we put the problem in a finite periodic volume. This explicitly
breaks rotation invariance, but leaves translation invariance intact (at least 
at the classical level), and leads to an energy spectrum with finite 
degeneracy. In order to clarify some subtle symmetry properties, we also
discuss issues of Hermiticity versus self-adjointness of various operators.

\subsection{Constant Magnetic Field on a Torus}

In this subsection we impose a torus boundary condition over a rectangular
region of size $L_x \times L_y$. This will lead to a quantization condition for
the magnetic flux. Since the magnetic field is constant, it obviously is 
periodic. On the other hand, the vector potential of the infinite volume theory
$A_x(x,y) = 0$, $A_y(x,y) = B x$ obeys the conditions
\begin{eqnarray}
&&A_x(x + L_x,y) = A_x(x,y), \nonumber \\
&&A_y(x + L_x,y) = A_y(x,y) + B L_x = A_y(x,y) + \p_y (B L_x y), \nonumber \\
&&A_x(x,y + L_y) = A_x(x,y), \nonumber \\
&&A_y(x,y + L_y) = A_y(x,y).
\end{eqnarray}
As a gauge-dependent quantity, the vector potential is periodic only up to
gauge transformations, i.e.
\begin{equation}
A_i(x + L_x,y) = A_i(x,y) - \p_i \varphi_x(y), \
A_i(x,y + L_y) = A_i(x,y) - \p_i \varphi_y(x).
\end{equation}
The gauge transformations $\varphi_x(y)$ and $\varphi_y(x)$ are transition
functions in a fiber bundle which specify the boundary condition. In our case
the transition functions are given by
\begin{equation}
\varphi_x(y) = \frac{\theta_x}{e} - B L_x y, \ 
\varphi_y(x) = \frac{\theta_y}{e}.
\end{equation}

Besides the field strength, gauge theories on a periodic volume possess 
additional gauge invariant quantities --- the so-called Polyakov loops --- 
which arise due to the non-trivial holonomies of the torus. For an Abelian 
gauge theory the Polyakov loops are defined as
\begin{equation}
\Phi_x(y) = \int_0^{L_x} dx \ A_x(x,y) - \varphi_x(y), \
\Phi_y(x) = \int_0^{L_y} dy \ A_y(x,y) - \varphi_y(x).
\end{equation}
In our case, they are given by
\begin{equation}
\Phi_x(y) = B L_x y - \frac{\theta_x}{e}, \
\Phi_y(x) = B L_y x - \frac{\theta_y}{e}.
\end{equation}

In order to respect gauge invariance of the theory on the torus, under shifts 
the wave function must also be gauge transformed accordingly
\begin{eqnarray}
\label{boundary}
&&\Psi(x + L_x,y) = \exp\left(i e \varphi_x(y)\right) \Psi(x,y) =
\exp\left(i \theta_x - i e B L_x y\right) \Psi(x,y), 
\nonumber \\
&&\Psi(x,y + L_y) = \exp\left(i e \varphi_y(x)\right) \Psi(x,y) =
\exp\left(i \theta_y\right) \Psi(x,y).
\end{eqnarray}
The angles $\theta_x$ and $\theta_y$ parametrize a family of self-adjoint
extensions of the Hamiltonian on the torus. Applying the boundary conditions 
from above in two different orders one obtains
\begin{eqnarray}
\Psi(x + L_x,y + L_y)&=& 
\exp\left(i \theta_x - i e B L_x (y + L_y)\right) 
\Psi(x,y + L_y) \nonumber \\
&=&\exp\left(i \theta_x + i \theta_y - 
i e B L_x (y + L_y)\right) \Psi(x,y), \nonumber \\
\Psi(x + L_x,y + L_y)&=&\exp\left(i \theta_y\right) \Psi(x + L_x,y)
\nonumber \\
&=&\exp\left(i \theta_x + i \theta_y - i e B L_x y \right) \Psi(x,y).
\end{eqnarray}
Hence, consistency of the boundary condition requires
\begin{equation}
\exp\left(- i e B L_x L_y \right) = 1 \ \Rightarrow \
B = \frac{2 \pi  n_\Phi}{e L_x L_y}, \ n_\Phi \in \Z.
\end{equation}
The total magnetic flux through the torus
\begin{equation}
\Phi = B L_x L_y = \frac{2 \pi n_\Phi}{e},
\end{equation}
is hence quantized in integer units of the elementary magnetic flux quantum 
$2 \pi/e$. Interestingly, the spectrum of a charged particle in a constant
magnetic field is discrete (but infinitely degenerate) already in the infinite
volume. As we will see, in the finite periodic volume it has only a finite 
$|n_\Phi|$-fold degeneracy determined by the number of flux quanta.

A quantum mechanical charged particle is sensitive to the complex phases
defined by the Polyakov loops
\begin{equation}
\exp(i e \Phi_x(y)) = \exp(i e B L_x y - i \theta_x), \
\exp(i e \Phi_y(x)) = \exp(i e B L_y x - i \theta_y),
\end{equation}
which are measurable in Aharonov-Bohm-type experiments. Remarkably, the 
Polya\-kov loops explicitly break the translation invariance of the torus at 
the quantum level. This is reminiscent of the quantum mechanical breaking of
CP invariance due to the $\theta$-vacuum angle in non-Abelian gauge theories.
The complex phases from above are invariant under shifts by integer multiples 
of
\begin{equation}
a_x = \frac{2 \pi}{e B L_y} = \frac{L_x}{n_\Phi}, \
a_y = \frac{2 \pi}{e B L_x} = \frac{L_y}{n_\Phi},
\end{equation}
in the $x$- and $y$-directions, respectively. Hence, at the quantum level the
continuous translation group of the torus is reduced to a discrete subgroup
which plays the role of the accidental symmetry group.

In this paper, we treat the gauge field as a classical background field, while
only the charged particle is treated quantum mechanically. It is interesting to
note that, once the gauge field is also quantized, the transition functions 
$\varphi_x(y)$ and $\varphi_y(x)$ become fluctuating physical degrees of
freedom of the gauge field. Still, as a consequence of
\begin{eqnarray}
A_i(x + L_x,y + L_y)&=&A_i(x,y + L_y) - \p_i \varphi_x(y + L_y) \nonumber \\
&=&A_i(x,y) - \p_i \varphi_x(y + L_y) - \p_i \varphi_y(x), \nonumber \\
A_i(x + L_x,y + L_y)&=&A_i(x + L_x,y) - \p_i \varphi_y(x + L_x) \nonumber \\
&=&A_i(x,y) - \p_i \varphi_y(x + L_x) - \p_i \varphi_x(y).
\end{eqnarray}
and of
\begin{eqnarray}
\Psi(x + L_x,y + L_y)&=&\exp(i e \varphi_x(y + L_y)) \Psi(x,y + L_y) 
\nonumber \\
&=&\exp(i e \varphi_x(y + L_y) + i e \varphi_y(x)) \Psi(x,y), \nonumber \\
\Psi(x + L_x,y + L_y)&=&\exp(i e \varphi_y(x + L_x)) \Psi(x + L_y,y) 
\nonumber \\
&=&\exp(i e \varphi_y(x + L_x) + i e \varphi_x(y)) \Psi(x,y),
\end{eqnarray}
the transition functions must obey the cocycle consistency condition
\begin{equation}
\varphi_y(x + L_x) + \varphi_x(y) - \varphi_x(y + L_y) - \varphi_y(x) = 
\frac{2 \pi n_\Phi}{e}.
\end{equation}
In this case, the magnetic flux $n_\Phi$ specifies a super-selection sector of
the theory. Analogous to the $\Z(N)^d$ center symmetry of non-Abelian $SU(N)$
gauge theories on a $d$-dimensional torus \cite{tHo79,tHo81}, Abelian gauge 
theories coupled to charged matter have a global $\Z^d$ center symmetry. The 
self-adjoint extension parameters $\theta_x$ and $\theta_y$ then turn into
conserved quantities (analogous to Bloch momenta) of the global $\Z^2$
symmetry on the 2-dimensional torus. In this sense, $\theta_x$ and $\theta_y$
are analogous to the $\theta$-vacuum angle of non-Abelian gauge theories, which
also distinguishes different super-selection sectors of the theory. The 
$\theta$-vacuum angle is a quantum mechanical source of explicit CP violation.
At the classical level, on the other hand, CP invariance remains intact because
$\theta$ does not affect the classical equations of motion. Similarly, for a 
charged particle on the torus the angles $\theta_x$ and $\theta_y$ characterize
the explicit breaking of continuous translation invariance down to a discrete
subgroup. Just like CP invariance for a non-Abelian gauge theory, for a charged
particle on the torus the full continuous translation symmetry remains intact
at the classical level, because $\theta_x$ and $\theta_y$ do not appear in the 
classical equations of motion.

In this paper, we treat the charged particle as a test charge which does not
surround itself with its own Coulomb field. This would change, once one would 
derive the charged particle from its own quantum field. For example, if one 
considers full-fledged QED, a single electron cannot even exist on the torus 
because the Coulomb field that surrounds it is incompatible with periodic
boundary conditions. Indeed, as a consequence of the Gauss law, the total 
charge on a torus always vanishes. To cure this problem, one could compensate
the charge of the electron by a classical background charge homogeneously
spread out over the torus. In our present calculation this is not necessary,
because the charged particle is treated as a test charge without its own 
surrounding Coulomb field. 

\subsection{Discrete Magnetic Translation Group}

As we have seen, in order to respect gauge invariance, on the torus the wave 
function must obey eq.(\ref{boundary}), which can be re-expressed as
\begin{equation}
\label{condition}
\Psi(x + L_x,y) = \exp\left(i \theta_x - \frac{2 \pi i n_\Phi y}{L_y}\right) 
\Psi(x,y), \ \Psi(x,y + L_y) = \exp(i \theta_y) \Psi(x,y).
\end{equation}
It is interesting to note that a factorization ansatz for the wave function as
in eq.(\ref{factorization}) is inconsistent with the boundary condition. Let us
consider the unitary shift operator generating translations by a distance $a_y$
in the $y$-direction as well as a $\theta_y$-dependent phase-shift
\begin{equation}
T_y = \exp\left(i P_y a_y - \frac{i \theta_y}{n_\Phi}\right) =
\exp\left(i \frac{P_y L_y - \theta_y}{n_\Phi}\right),
\end{equation}
which acts as
\begin{equation}
T_y \Psi(x,y) = \exp\left(- \frac{i \theta_y}{n_\Phi}\right) \Psi(x,y + a_y).
\end{equation}
Obviously, $T_y$ commutes with the Hamiltonian because $P_y$ does. Indeed, the
shifted wave function does obey the boundary condition eq.(\ref{condition}), 
i.e.
\begin{eqnarray}
T_y \Psi(x + L_x,y)&=&
\exp\left(- \frac{i \theta_y}{n_\Phi}\right) \Psi(x + L_x,y + a_y) \nonumber \\
&=&\exp\left(- \frac{i \theta_y}{n_\Phi}\right) 
\exp\left(i \theta_x - \frac{2 \pi i n_\Phi (y + a_y)}{L_y}\right) 
\Psi(x,y + a_y) 
\nonumber \\
&=&\exp\left(i \theta_x - \frac{2 \pi i n_\Phi y}{L_y}\right) T_y \Psi(x,y),
\end{eqnarray}
which is the case only because
\begin{equation}
a_y = \frac{L_y}{n_\Phi} \ \Rightarrow \
\exp\left(- \frac{2 \pi i n_\Phi a_y}{L_y}\right) = 1.
\end{equation}
Furthermore, we also have
\begin{eqnarray}
T_y \Psi(x,y + L_y)&=&
\exp\left(- \frac{i \theta_y}{n_\Phi}\right) \Psi(x,y + a_y + L_y) \nonumber \\
&=&\exp\left(i \theta_y - \frac{i \theta_y}{n_\Phi}\right) \Psi(x,y + a_y) =
\exp(i \theta_y) T_y \Psi(x,y).
\end{eqnarray}
Hence, as we argued before, the translations in the $y$-direction are reduced 
to the discrete group $\Z(n_\Phi)$. In particular, all translations $T_y^{n_y}$
compatible with the boundary conditions can be expressed as the $n_y$-th power
of the elementary translation $T_y$. According to eq.(\ref{RP}), 
$P_y = - e B R_x$, such that
\begin{equation}
T_y = \exp\left(i \frac{L_y P_y - \theta_y}{n_\Phi}\right) = 
\exp\left(- i \frac{e B L_y R_x + \theta_y}{n_\Phi}\right) =
\exp\left(- i \left(\frac{2 \pi i R_x}{L_x} + 
\frac{\theta_y}{n_\Phi}\right)\right).
\end{equation} 
Similarly, up to gauge transformations the operator 
\begin{eqnarray}
T_x&=&\exp\left(i P_x a_x - \frac{i \theta_x}{n_\Phi}\right) = 
\exp\left(i e B R_y a_x - \frac{i \theta_x}{n_\Phi}\right) \nonumber \\
&=&\exp\left(\frac{2 \pi i n_\Phi R_y a_x}{L_x L_y} -
\frac{i \theta_x}{n_\Phi}\right) = 
\exp\left(i \left(\frac{2 \pi i R_y}{L_y} -
\frac{\theta_x}{n_\Phi}\right)\right)
\end{eqnarray}
generates translations in the $x$-direction.
Since on the torus the Runge-Lenz vector component $R_y$, which determines the 
$y$-coordinate of the center of the cyclotron orbit, is defined only modulo
$L_y$, it is indeed natural to consider the translation operator $T_x$. In 
fact, although it formally commutes with the Hamiltonian, the operator $R_y$ 
itself is no longer self-adjoint in the Hilbert space of wave functions on the 
torus. The operator $T_x$, on the other hand, does act as a unitary operator in
the Hilbert space. It is worth noting that, at least in the gauge we have 
picked, the operator $R_x$ is still self-adjoint. However, this would not be 
the case, for example, in the symmetric gauge, and it is hence most natural to
work with $T_x$ and $T_y$ instead of $R_x$ and $R_y$ or equivalently $P_x$ and
$P_y$.

The boundary condition of eq.(\ref{condition}) can now be expressed as
\begin{equation}
\label{conditionT}
T_x^{n_\Phi} \Psi(x,y) = \Psi(x,y), \ T_y^{n_\Phi} \Psi(x,y) = \Psi(x,y).
\end{equation}
As a consequence of the commutation relation $[R_x,R_y] = i/e B$, one obtains
\begin{equation}
\label{Tcom}
T_y T_x = \exp\left(\frac{2 \pi i}{n_\Phi}\right) T_x T_y.
\end{equation}
This implies that
\begin{equation}
T_x \Psi(x,y) = \exp\left(\frac{2 \pi i y}{L_y} - \frac{i \theta_x}{n_\Phi}
\right) \Psi(x + \frac{L_x}{n_\Phi},y),
\end{equation}
i.e., up to a periodic gauge transformation 
$\exp(2 \pi i y/L_y - i \theta_x/n_\Phi)$, $T_x$ translates the wave function 
by a distance $L_x/n_\Phi$.

Remarkably, although at the classical level the torus has two continuous 
translation symmetries, the corresponding infinitesimal generators $P_x$ and 
$P_y$ are not self-adjoint in the Hilbert space of wave functions on the torus.
Only the finite translations $T_x$ and $T_y$ are represented by unitary 
operators, which, however, do not commute with each other. The two operators 
$T_x$ and $T_y$ generate a discrete translation group ${\cal G}$ consisting of 
the elements
\begin{eqnarray}
&&g(n_x,n_y,m) = \exp\left(\frac{2 \pi i m}{n_\Phi}\right) T_y^{n_y} T_x^{n_x},
\nonumber \\
&&n_x, n_y, m \in \{0,1,2,...,n_\Phi - 1\}.
\end{eqnarray}
The group multiplication rule takes the form
\begin{equation}
g(n_x,n_y,m) g(n_x',n_y',m') = 
g(n_x + n_x',n_y + n_y',m + m' - n_x n_y'),
\end{equation}
with all summations being understood modulo $n_\Phi$. Obviously, the unit 
element is represented by
\begin{equation}
\1 = g(0,0,0),
\end{equation}
while the elements
\begin{equation}
z_m = g(0,0,m) = \exp\left(\frac{2 \pi i m}{n_\Phi}\right),
\end{equation}
form the cyclic Abelian subgroup $\Z(n_\Phi) \subset {\cal G}$. The inverse of 
a general group element $g(n_x,n_y,m)$ is given by
\begin{equation}
g(n_x,n_y,m)^{-1} = g(- n_x,- n_y,- m - n_x n_y),
\end{equation}
because
\begin{equation}
g(n_x,n_y,m) g(- n_x,- n_y,- m - n_x n_y) = g(0,0,- n_x n_y + n_x n_y) = 
g(0,0,0) = \1.
\end{equation}
It is interesting to consider the conjugacy class of a group element 
$g(n_x,n_y,m)$ which consists of the elements
\begin{eqnarray}
&&\!\!\!\!\!\!\!\!\!\!g(n_x',n_y',m') g(n_x,n_y,m) g(n_x',n_y',m')^{-1} = 
\nonumber \\
&&\!\!\!\!\!\!\!\!\!\!g(n_x' + n_x,n_y' + n_y,m' + m - n_x' n_y) 
g(- n_x',- n_y',- m' - n_x' n_y') = \nonumber \\
&&\!\!\!\!\!\!\!\!\!\!g(n_x,n_y,m - n_x'(n_y + n_y') + (n_x' + n_x) n_y') =
g(n_x,n_y,m + n_x n_y' - n_x' n_y).
\end{eqnarray}
In particular, as one would expect, the elements $g(0,0,m) = z_m \in \Z(n_\Phi)$
are conjugate only to themselves and thus form $n_\Phi$ single-element 
conjugacy classes. Obviously, multiplication by a phase $z_m$ is just a global 
gauge transformation and thus leaves the physical state invariant. Hence, the 
conjugacy classes correspond to gauge equivalence classes. 

The elements $g(0,0,m) = z_m$ commute with all other elements and thus form the 
center $\Z(n_\Phi)$ of the group ${\cal G}$. Since the individual elements of 
the center form separate conjugacy classes, the center is a normal subgroup
and can hence be factored out. The center itself represents global phase
transformations of the wave function, and hence factoring it out corresponds
to identifying gauge equivalence classes. Physically speaking, the quotient 
space ${\cal G}/\Z(n_\Phi) = \Z(n_\Phi) \times \Z(n_\Phi)$ corresponds to 
discrete translations up to gauge transformations. It should be pointed out 
that ${\cal G}$ is not simply given by the direct product 
$\Z(n_\Phi) \times \Z(n_\Phi) \times \Z(n_\Phi)$. In fact, the quotient space
$\Z(n_\Phi) \times \Z(n_\Phi)$ is not a subgroup of ${\cal G}$, and hence 
${\cal G}$ is also not the semi-direct product of $\Z(n_\Phi) \times \Z(n_\Phi)$
and $\Z(n_\Phi)$. All we can say (besides defining the group ${\cal G}$ as done
before) is that it is a particular central extension of 
$\Z(n_\Phi) \times \Z(n_\Phi)$ by the center subgroup $\Z(n_\Phi)$.

\subsection{Spectrum and Degeneracy on the Torus}

Let us first discuss the classical problem on the torus. In that case, the
magnetic flux need not be quantized. Also the values of the Polyakov loop are
not detectable at the classical level because they have no effect on the motion
of a test charge, which is entirely determined by the Lorentz force. The
classical orbits of a charged particle in a constant magnetic field on the 
torus still are closed circles. However, as illustrated in figure 1, the circle
may close only after wrapping around the periodic boundary.
\begin{figure}[htb]
\begin{center}
\epsfig{file=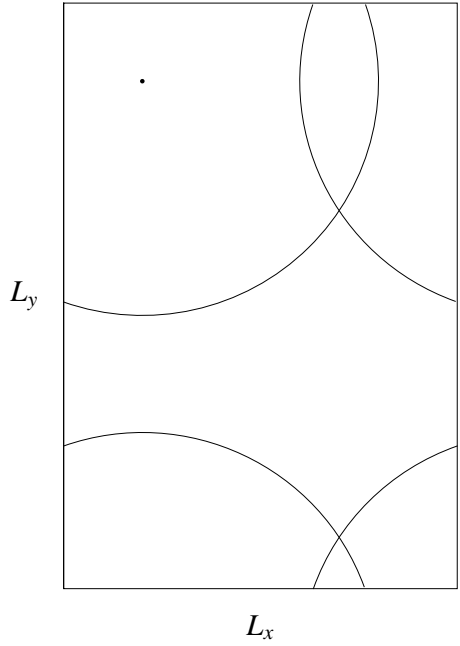,width=6cm}
\end{center}
\caption{\it Closed cyclotron orbit of a charged particle in a constant 
magnetic field on the torus. The circular orbit closes after wrapping around
the periodic boundary several times. The dot marks the center of the circle.}
\end{figure}
Since all classical orbits are still closed, one expects that on the torus
the accidental symmetry is still present. 

It should be pointed out that on the torus the Hamiltonian is identically the 
same as in the infinite volume. It now just acts on the restricted set of wave 
functions obeying the boundary condition eq.(\ref{condition}). In particular, 
the finite volume wave functions are appropriate linear combinations of the 
infinitely many degenerate states of a given Landau level. As a result, the 
energy spectrum remains unchanged, but the degeneracy is substantially reduced.

Let us use the fact that $T_y$ commutes with the Hamiltonian to construct
simultaneous eigenstates of both $H$ and $T_y$. Since for states on the torus 
$T_y^{n_\Phi} = \1$, the eigenvalues of $T_y$ are given by 
$\exp(2 \pi i l_y/n_\phi)$ with $l_y \in \{0,1,...,n_\Phi - 1\}$, while the 
eigenvalues of $H$ are still given by $E_n = \omega(n + \frac{1}{2})$. Hence, 
we can construct simultaneous eigenstates $|n l_y\rangle$ such that
\begin{equation}
H |n l_y\rangle = \omega \left(n + \frac{1}{2}\right) |n l_y\rangle, \
T_y |n l_y\rangle = \exp\left(\frac{2 \pi i l_y}{n_\Phi}\right) 
|n l_y\rangle.
\end{equation}
The states $|n l_y\rangle$ are the finite-volume analog of the states 
$|n p_y \rangle$ of eq.(\ref{eigenstates}) with 
$p_y = (2 \pi l_y + \theta_y)/L_y$. In
coordinate representation these states are given by the wave functions
\begin{eqnarray}
\langle \vec x|n l_y\rangle&=&A \sum_{n_x \in \Z} 
\psi_n\left(x + \left(n_\Phi n_x + l_y + \frac{\theta_y}{2 \pi}\right) 
\frac{L_x}{n_\Phi}\right) \nonumber \\
&\times&
\exp\left(\frac{2 \pi i y}{L_y}\left(n_\Phi n_x + l_y + \frac{\theta_y}{2 \pi}
\right) - i \theta_x n_x\right).
\end{eqnarray}
As a special case, let us consider the ground state for $n_\Phi = 1$, which is
non-degenerate
\begin{eqnarray}
\langle \vec x|n = 0,l_y = 0\rangle&=&A \sum_{n_x \in \Z} 
\psi_0\left(x + \left(n_x + \frac{\theta_y}{2 \pi}\right) L_x\right) 
\nonumber \\
&\times&\exp\left(\frac{2 \pi i y}{L_y}\left(n_x + \frac{\theta_y}{2 \pi}
\right) - i \theta_x n_x\right).
\end{eqnarray}
\begin{figure}[htb]
\begin{center}
\epsfig{file=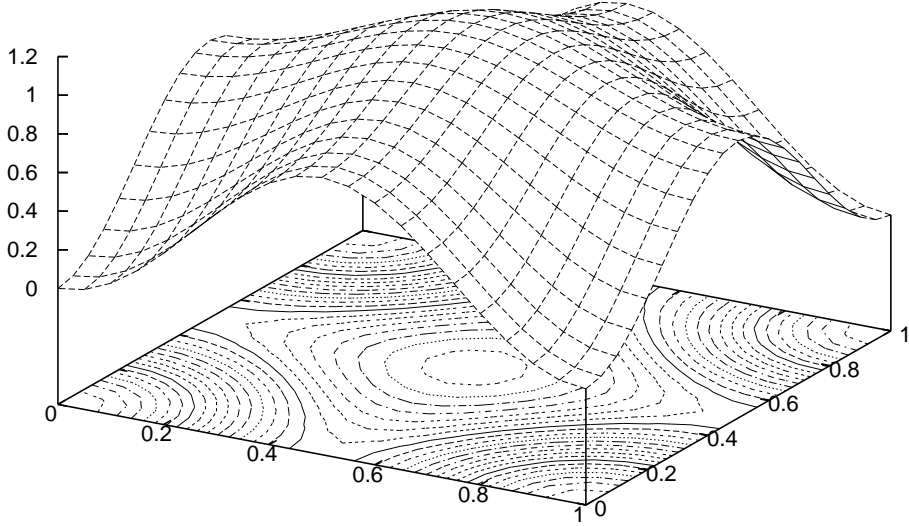,width=15cm}
\end{center}
\caption{\it Probability density for the state 
$\langle \vec x|n = 0,l_y = 0\rangle$ with $\theta_x = \theta_y = \pi$ and
$n_\Phi = 1$ over a square-shaped torus with $M L_x = M L_y = 1$.}
\end{figure}
In this state the probability density, which is illustrated in figure 2, has 
its maximum at $(- L_x \theta_y/2 \pi,L_y \theta_x/2 \pi)$. This shows once 
again that the self-adjoint extension parameters $\theta_x$ and $\theta_y$ 
indeed explicitly break translation invariance.

As a consequence of eq.(\ref{Tcom}) one obtains
\begin{equation}
T_y T_x |n l_y\rangle = \exp\left(\frac{2 \pi i}{n_\phi}\right) T_x T_y
|n l_y\rangle = \exp\left(\frac{2 \pi i (l_y + 1)}{n_\Phi}\right) 
T_x |n l_y\rangle,
\end{equation}
from which we conclude that
\begin{equation}
T_x |n l_y\rangle = |n (l_y + 1)\rangle.
\end{equation}
Since $[T_x,H] = 0$, the $n_\Phi$ states $|n l_y\rangle$ with 
$l_y \in 0,1,...,n_\Phi - 1$ thus form an irreducible representation of the
magnetic translation group. Using $n_\Phi = 4$ as a concrete example, a matrix
representation of the two generators of ${\cal G}$ is given by
\begin{equation}
T_x = \left(\begin{array}{cccc} 0 & 0 & 0 & 1 \\ 1 & 0 & 0 & 0 \\
0 & 1 & 0 & 0 \\ 0 & 0 & 1 & 0 \end{array}\right), \
T_y = \left(\begin{array}{cccc} 1 & 0 & 0 & 0 \\ 
0 & i & 0 & 0 \\ 0 & 0 & - 1 & 0 \\ 0 & 0 & 0 & - i \end{array} \right).
\end{equation}

Similarly, one can construct simultaneous eigenstates $|n l_x\rangle$ of $H$ 
and $T_x$
\begin{equation}
H |n l_x\rangle = \omega \left(n + \frac{1}{2}\right) |n l_x\rangle, \
T_x |n l_x\rangle = \exp\left(\frac{2 \pi i l_x}{n_\phi}\right) 
|n l_x\rangle.
\end{equation}
The states $|n l_x\rangle$ are the finite-volume analog of the states 
$|n p_x \rangle$ of eq.(\ref{eigenx}) with $p_x = (2 \pi l_x + \theta_x)/L_x$.
In coordinate representation these states are given by the wave functions
\begin{eqnarray}
\langle \vec x|n l_x\rangle&=&A \sum_{n_y \in \Z} 
\psi_n\left(y - \left(n_\Phi n_y + l_x + \frac{\theta_x}{2 \pi}\right) 
\frac{L_y}{n_\Phi}\right) \nonumber \\
&\times&
\exp\left(\frac{2 \pi i x}{L_x}\left(n_\Phi n_y + l_x + \frac{\theta_x}{2 \pi} -
\frac{n_\Phi y}{L_y}\right) + i \theta_y n_y\right).
\end{eqnarray}
It is worth noting that
\begin{equation}
T_y |n l_x\rangle = |n (l_x - 1)\rangle.
\end{equation}
Similar to the infinite volume case, it is straightforward to show that the two
sets of eigenstates $\langle \vec x|n l_y\rangle$ and 
$\langle \vec x|n l_x\rangle$ span the same subspace of the Hilbert space.
In particular, for $n_\Phi = 1$ the ground state is non-degenerate and one can
show that
\begin{equation}
|n = 0, l_x = 0\rangle = |n = 0, l_y = 0\rangle.
\end{equation}

As we have seen, on the torus continuous translation invariance is explicitly 
broken down to the discrete magnetic translation group ${\cal G}$ by the 
self-adjoint extension parameters $\theta_x$ and $\theta_y$. Still, all states 
(including the ground state) remain degenerate. However, unlike in the infinite
volume, the degeneracy is reduced to a finite amount $n_\Phi$. Only when one 
varies $\theta_x$ and $\theta_y$ one recovers the infinite degeneracy of the
infinite system. As in the infinite volume, one may ask if the degenerate 
ground state indicates that the discrete magnetic translation group ${\cal G}$
is spontaneously broken. While there are striking similarities with spontaneous
symmetry breaking, there are also important differences. First of all, when a
system with a broken symmetry is put in a finite volume, the symmetry is 
usually restored dynamically. For example, this is the case for the 
spontaneously broken $SU(2)_L \otimes SU(2)_R$ chiral symmetry in QCD 
as well as for the spontaneously broken $SU(2)_s$ spin symmetry in
antiferromagnets. An important exception are ferromagnets for which the ground
state remains exactly degenerate even in a finite volume. This is a consequence
of the fact that the magnetization order parameter of a ferromagnet is a 
conserved quantity, while the staggered magnetization order parameter of an 
antiferromagnet is not conserved. In this sense, the charged particle in a
magnetic field behaves like a ferromagnet. The ``order parameter'' that signals
the ``spontaneous breakdown'' of translation invariance is the Runge-Lenz 
vector $(R_x,R_y)$ pointing to the center of the cyclotron circle, which is 
indeed a conserved quantity.

\subsection{Coherent States on the Torus}

It is interesting to construct coherent states $|\lambda \lambda'\rangle_T$
for the particle on the torus. This is achieved by superposition of shifted 
copies of the coherent state $|\lambda \lambda'\rangle$ of the system in the 
infinite volume
\begin{equation}
|\lambda \lambda'\rangle_T = A \sum_{n_x, n_y \in \Z} 
T_x^{n_\Phi n_x} T_y^{n_\Phi n_y} |\lambda \lambda'\rangle.
\end{equation}
By construction, this state obeys the boundary condition eq.(\ref{conditionT}).
The factor $A$ is determined from the normalization condition 
\begin{eqnarray}
_T\langle\lambda \lambda'|\lambda \lambda'\rangle_T&=&
|A|^2 \sum_{m_x,m_y \in \Z} (-1)^{n_\Phi m_x m_y} 
\exp\left(- \frac{\pi^2}{M \omega} \left(\frac{n_\Phi^2 m_x^2}{L_y^2}
+ \frac{n_\Phi^2 m_y^2}{L_x^2} \right)\right) \nonumber \\
&\times&
\exp\left(i \left(\frac{2 \pi \langle R_y \rangle}{L_y} - 
\frac{\theta_x}{n_\Phi}\right) n_\Phi m_x\right)
\exp\left(- i \left(\frac{2 \pi \langle R_x \rangle}{L_y} +
\frac{\theta_y}{n_\Phi}\right) n_\Phi m_y \right) \nonumber \\ 
&=& 1.
\end{eqnarray}
It is easy to see that a finite-volume coherent state remains coherent during 
the time-evolution. Just as in the infinite volume, $\lambda(t) =
\lambda(0) \exp(- i \omega t)$, while $\lambda'$ is time-independent. In the
infinite volume $\lambda' = \sqrt{M \omega/2} (\langle R_x \rangle + i 
\langle R_y \rangle)$ determines the position of the center of the cyclotron 
circle. On the torus, this center is well-defined only up to shifts by 
multiples of $L_x$ or $L_y$. Indeed one finds
\begin{eqnarray}
\label{expectT}
&&_T\langle\lambda \lambda'|T_x^{l_x}|\lambda \lambda'\rangle_T = B_{l_x}
\exp\left(i \left(\frac{2 \pi \langle R_y \rangle}{L_y} -
\frac{\theta_x}{n_\Phi}\right) l_x\right), \nonumber \\
&&_T\langle\lambda \lambda'|T_y^{l_y}|\lambda \lambda'\rangle_T = B_{l_y}
\exp\left(- i \left(2 \pi \frac{\langle R_x \rangle}{L_x} +
\frac{\theta_y}{n_\Phi}\right) l_y\right),
\end{eqnarray}
which shows that (together with $\theta_x$ and $\theta_y$) the expectation
values $\langle R_x \rangle$ and $\langle R_y \rangle$ of the infinite volume
coherent state determine the position of the center of the cyclotron circle
(the Runge-Lenz vector) modulo the periodicity lengths $L_x$ and $L_y$ of the
torus. The prefactors in eq.(\ref{expectT}) take the form
\begin{eqnarray}
B_{l_x}&=&|A|^2 \sum_{m_x,m_y \in \Z} (-1)^{(n_\Phi m_x + l_x) m_y} 
\exp\left(- \frac{\pi^2}{M \omega} \left(\frac{(n_\Phi m_x + l_x)^2}{L_y^2}
+ \frac{n_\Phi^2 m_y^2}{L_x^2} \right)\right) \nonumber \\
&\times&\exp\left(i \left(\frac{2 \pi \langle R_y \rangle}{L_y} -
\frac{\theta_x}{n_\Phi}\right) n_\Phi m_x \right)
\exp\left(- i \left(\frac{2 \pi \langle R_x \rangle}{L_y} +
\frac{\theta_y}{n_\Phi}\right) n_\Phi m_y \right), \nonumber \\
B_{l_y}&=&|A|^2 \sum_{m_x,m_y \in \Z} (-1)^{(n_\Phi m_y + l_y) m_x} 
\exp\left(- \frac{\pi^2}{M \omega} \left(\frac{n_\Phi^2 m_x^2}{L_y^2}
+ \frac{(n_\Phi m_y + l_y)^2}{L_x^2} \right)\right) \nonumber \\
&\times&\exp\left(i \left(\frac{2 \pi \langle R_y \rangle}{L_y} -
\frac{\theta_x}{n_\Phi}\right) n_\Phi m_x \right)
\exp\left(- i \left(\frac{2 \pi \langle R_x \rangle}{L_y} +
\frac{\theta_y}{n_\Phi}\right) n_\Phi m_y \right). \nonumber \\ \,
\end{eqnarray}
Finally, let us consider the coherent states with $\lambda = 0$ but arbitrary
$\lambda'$. Just as in the infinite volume, these states are ground states with
minimal energy $\omega/2$. Indeed, for $n_\Phi = 1$ (i.e.\ when there is no
degeneracy) one can show that
\begin{equation}
|\lambda = 0,\lambda'\rangle = |n = 0,l_x = 0\rangle = |n = 0,l_y = 0\rangle,
\end{equation}
(provided that the arbitrary complex phase of $|\lambda = 0,\lambda'\rangle$ is
chosen appropriately).

\section{Conclusions}

We have re-investigated an old and rather well-studied problem in quantum 
mechanics --- a charged particle in a constant magnetic field --- from an
unconventional accidental symmetry perspective. The fact that all
classical cyclotron orbits are closed circles identifies the center of the
circle as a conserved quantity analogous to the Runge-Lenz vector of the Kepler
problem. Remarkably, (up to gauge transformations) the corresponding 
``accidental'' symmetry is just translation invariance. In particular, the 
coordinates $(R_x,R_y) = (- P_y,P_x)/eB$ of the center of the cyclotron circle 
simultaneously generate infinitesimal translations $- P_y$ and $P_x$ (up to 
gauge transformations) in the $y$- and $x$-directions, respectively. As is 
well-known, in a constant magnetic field translations in the $x$- and 
$y$-directions do not commute, i.e.\ $[P_x,P_y] = i e B$, and thus the two 
coordinates $R_x$ and $R_y$ of the center of the cyclotron circle are also not 
simultaneously measurable at the quantum level. In contrast, the radius of the
cyclotron circle has a sharp value in an energy eigenstate. The accidental 
symmetry leads to the infinite degeneracy of the Landau levels. 

In order to further investigate the nature of the accidental symmetry, we have 
put the system in a finite rectangular periodic volume. Obviously, this breaks 
rotation invariance, but leaves translation invariance (and thus the accidental
symmetry) intact --- at least at the classical level. Interestingly, at the
quantum level continuous translation invariance is explicitly broken down to a
discrete magnetic translation group, due to the existence of two angles 
$\theta_x$ and $\theta_y$ which parametrize a family of self-adjoint extensions
of the Hamiltonian on the torus. In a field theoretical context, in which the
gauge field is dynamical (and not just treated as a classical background 
field), the parameters $\theta_x$ and $\theta_y$ characterize super-selection
sectors. In this sense, they are analogous to the vacuum angle $\theta$ of 
non-Abelian gauge theories. Just as the $\theta$-vacuum angle explicitly
breaks CP invariance at the quantum level but is classically invisible, the 
angles $\theta_x$ and $\theta_y$ lead to a quantum mechanical explicit breaking 
of continuous translation invariance down to the discrete magnetic translation 
group. The magnetic translation group ${\cal G}$ itself, which plays the role
of the accidental symmetry in the periodic volume, is a particular central 
extension of $\Z(n_\Phi) \otimes \Z(n_\Phi)$ by the center subgroup 
$\Z(n_\Phi)$, where $n_\Phi$ is the number of magnetic flux quanta trapped in 
the torus. We find it remarkable that the simple fact that all classical
cyclotron orbits are closed circles has such intricate effects at the quantum
level.

We have also discussed the relation of ground state degeneracy with the 
possible spontaneous breakdown of translation invariance. Indeed the Runge-Lenz
vector (which points to the center of the cyclotron orbit) acts as a 
corresponding ``order parameter''. Just like the magnetization in a ferromagnet
(but unlike the staggered magnetization in an antiferromagnet), the Runge-Lenz 
vector is a conserved quantity. Consequently, the ground state remains 
degenerate even in a finite volume. Furthermore, just as the three components 
of the magnetization vector do not commute with each other, the two components 
of the Runge-Lenz vector are also not simultaneously measurable. Still, unlike 
a ferromagnet, a single charged particle in a magnetic field has just a finite 
number of degrees of freedom and can thus not display all features usually 
associated with spontaneous symmetry breaking. In particular, in the system 
discussed in this paper there is no room for massless Goldstone excitations.

While many aspects of the Landau level problem are well-known, we hope that we 
have painted a picture of cyclotron motion that reveals new aspects of this 
fascinating system, which behaves in a unique and sometimes counter-intuitive 
manner. 

\section*{Acknowledgments}

We like to thank Nathan Habegger for an illuminating discussion about the 
theory of discrete groups. This work is supported in parts by the 
Schweizerischer Na\-tio\-nal\-fonds.


\begin{thebibliography}{10}

\bibitem{Foc35}
V.\ Fock, Z.\ Physik 98 (1935) 145.

\bibitem{Bar36}
V.\ Bargmann, Z.\ Physik 99 (1936) 576.

\bibitem{Len24}
W.\ Lenz, Z.\ Physik 24 (1924) 197.

\bibitem{McI71}
H.\ V.\ McIntosh, in Group Theory and its Applications, Vol.\ 2 (1971) 75, 
Academic Press, Inc.\, New York and London.

\bibitem{Has07}
M.\ H.\ Al-Hashimi and U.-J.\ Wiese, Ann.\ Phys.\ 323 (2008) 82.

\bibitem{Lan30}
L.\ D.\ Landau, Z.\ Physik 64 (1930) 629.

\bibitem{Joh49}
M.\ H.\ Johnson and B.\ A.\ Lippmann, Phys.\ Rev.\ 76 (1949) 828.

\bibitem{Dul66}
V.\ A.\ Dulock and H.\ V.\ McIntosh, J.\ Math.\ Phys.\ 7 (1966) 1401.

\bibitem{Zak64}
J.\ Zak, Phys.\ Rev.\ 134 (1964) A1602.

\bibitem{Ste08}
A.\ Stern, Ann.\ Phys.\ 323 (2008) 204.

\bibitem{Che95}
G.-H.\ Chen, L.-M.\ Kuang, and M.-L.\ Ge, Phys.\ Rev.\ B53 (1996) 9540.

\bibitem{Zai89}
H.\ Zainuddin, Phys.\ Rev.\ D40 (1989) 636.

\bibitem{Fel70}
A.\ Feldman and A.\ H.\ Kahn, Phys.\ Rev.\ B12 (1970) 4584.

\bibitem{tHo79}
G.\ 't Hooft, Nucl.\ Phys.\ B153 (1979) 141.

\bibitem{tHo81}
G.\ 't Hooft, Commun.\ Math.\ Phys 81 (1981) 267.

\end{thebibliography}
\end{document}